\documentclass[runningheads]{llncs}
\usepackage[T1]{fontenc}%
\usepackage{graphicx}
\usepackage{stmaryrd}
\usepackage{comment}
\usepackage{amsfonts}
\usepackage{hyperref}

\begin{document}
\title{Across-subject ensemble-learning alleviates the need for large samples for fMRI decoding}
\titlerunning{Across-subject ensemble-learning for fMRI decoding}
\author{Himanshu Aggarwal\inst{1}\orcidID{0000-0003-2451-0470} \and
Liza Al-Shikhley\inst{1}\orcidID{0009-0002-0297-4231} \and
Bertrand Thirion\inst{1}\orcidID{0000-0001-5018-7895}}
%
% index{Aggarwal, Himanshu}
% index{Al-Shikhley, Liza}
% index{Thirion, Bertrand}
%
\authorrunning{H. Aggarwal et al.}
\institute{Inria, CEA, Universit{\'e} Paris-Saclay,  Palaiseau, 91120, France \\
\email{himanshu.aggarwal@inria.fr}\\
\email{liza.alshikhley@gmail.com}\\
\email{bertrand.thirion@inria.fr}\\
\url{https://team.inria.fr/mind/}\\
}
\maketitle
\begin{abstract}
Decoding cognitive states from functional magnetic resonance imaging is central to understanding the functional organization of the brain. Within-subject decoding avoids between-subject correspondence problems but requires large sample sizes to make accurate predictions; obtaining such large sample sizes is both challenging and expensive. Here, we investigate an ensemble approach to decoding that combines the classifiers trained on data from other subjects to decode cognitive states in a new subject. We compare it with the conventional decoding approach on five different datasets and cognitive tasks. We find that it outperforms the conventional approach by up to 20\% in accuracy, especially for datasets with limited per-subject data. The ensemble approach is particularly advantageous when the classifier is trained in voxel space. Furthermore, a Multi-layer Perceptron turns out to be a good default choice as an ensemble method. These results show that the pre-training strategy reduces the need for large per-subject data.

\end{abstract}
\section{Introduction}
Brain \emph{decoding} refers to the process of inferring cognitive states from an individual's brain signals.
It is an important tool to understand how information processing is distributed in the brain \cite{Zhang2021} and to spot potential dysfunction in neurology or psychiatry \cite{Sarraf2017}.
It entails acquiring brain signals using neuroimaging techniques such as functional magnetic resonance imaging (fMRI) and then training a classifier to learn the mapping from brain signals to cognitive states.
However, given that acquiring such data is expensive and time-consuming, the number of features (voxels) is usually much larger than the number of samples (repetitions of stimulus presentation).
This high feature-to-sample ratio, together with low data signal-to-noise ratio is detrimental to the quality of predictions \cite{Haynes2015}, hence solutions to increase the amount of available data are needed.

\paragraph{Across-subject correspondence issues:}
One way to increase the sample size could simply be to collect data from many individuals and then train classifiers across these individuals to decode the brain activity of a new individual.
However, it has been shown that spatio-temporal patterns of brain activity generalize poorly across individuals \cite{Gratton2018} leading to sub-optimal decoding accuracy \cite{Shinkareva2008}.
Similar issues also exist in decoding brain signals from other neuroimaging techniques like magneto-/electro-encephalography (M/EEG) that have direct applications in brain-computer interfaces (BCI) \cite{Thompson2019}.
A common solution in M/EEG-BCI research is aimed towards efficiently using the available information \emph{within-subjects} by combining classifiers via voting or stacking \cite{Lotte2007,Lotte2018}.
The combination of classifiers can be done at the feature level or the decision level \cite{Lotte2018}.

\paragraph{Ensemble-learning:}
One recent study \cite{Gu2022} showed that fMRI-based brain activity \emph{encoding} benefits from a linear combination of other subjects' predicted response vectors.
%
\begin{comment}
Brain \emph{encoding} is the reverse of decoding, where the goal is to predict brain signals from cognitive states.
%
So as proposed in \cite{Gu2022}, if $r_j$ is the encoding model trained to predict the brain response vector of subject $j$ given a stimulus $S$.
%
Then subject $i$'s predicted response vector for stimulus $S$ is given by $w_{i,0} + \sum_{j=1}^{N-1} w_{ij} r_j(S)$, where $w_{i,0}$ is the intercept, $w_{ij}$ is the weight assigned to subject $j$ for subject $i$'s prediction and $N-1$ is the number of subjects except subject $i$.
\end{comment}
%
However, to our knowledge, this approach of combining models and learning from an \emph{ensemble} of individuals has not been investigated within the context of decoding, which is addressed in
the present study.

\section{Methods}

\begin{figure}[tb]
\centerline{\includegraphics[width=0.85\textwidth]{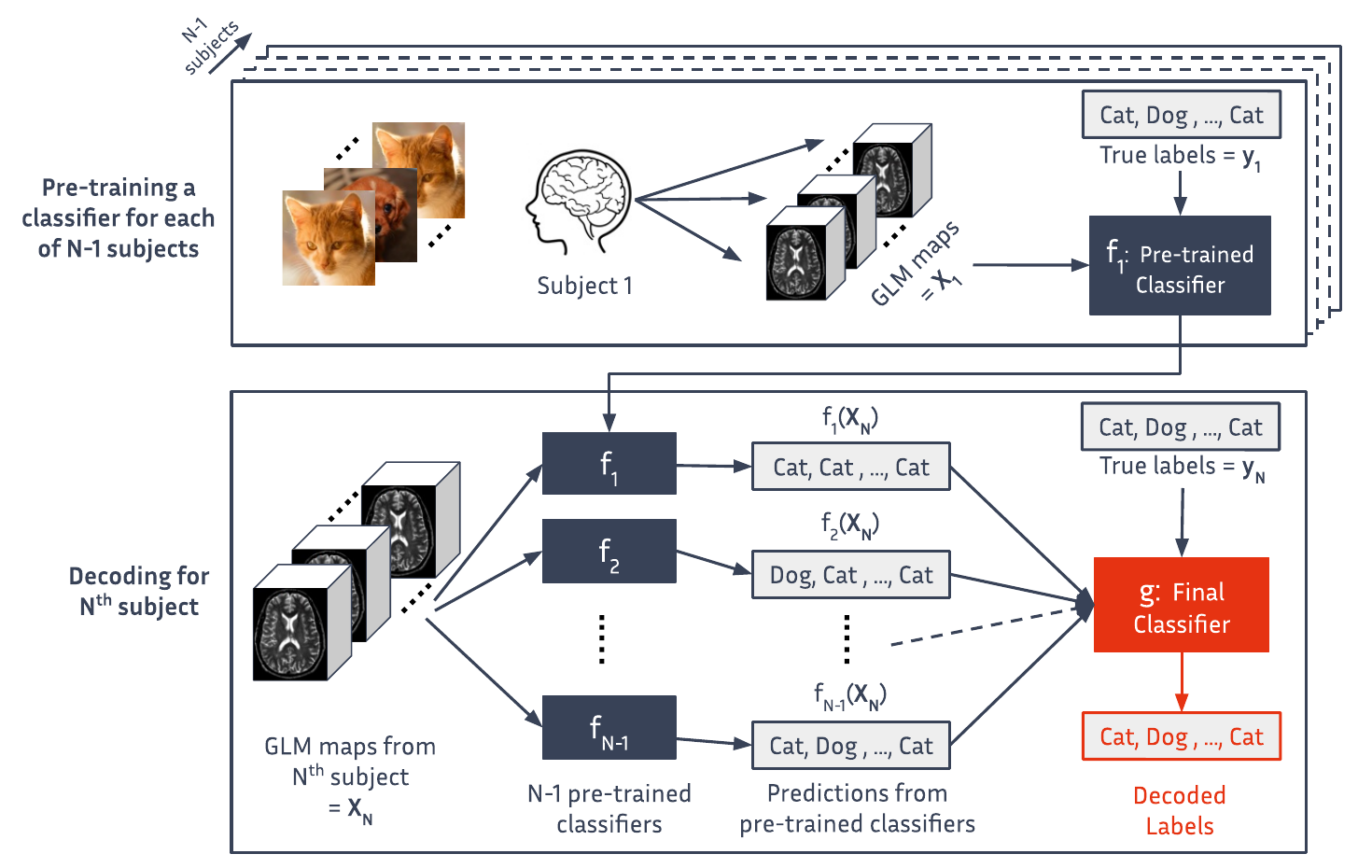}}
\caption{\textbf{Ensemble-learning (by stacking):} Ensemble-learning (by stacking) involves (top) pre-training separate classifiers for each of $N-1$ subjects and then (bottom) training a final classifier to learn the mapping between predictions from each pre-trained classifier and the true labels for $N^{th}$ subject.}
\label{figure:method}
\end{figure}

\paragraph{Ensemble-learning (by stacking):}
In a conventional decoding setting, a classifier is trained to learn the mapping between stimuli labels and a given feature space.
Let $\mathbf{X}_1,\cdots,\mathbf{X}_N  \in \mathcal{X}^N$ be the fMRI datasets acquired in $N$ subjects, together with the corresponding labels $\mathbf{y}_1,\cdots,\mathbf{y}_N \in \mathcal{Y}^N $. 
Here, we present an ensemble approach where we first \emph{(i)} pre-train separate classifiers $f_{i}$ for each subject $i \in \llbracket N - 1 \rrbracket$ to learn the mapping between their respective feature space $\mathcal{X}$ and stimuli labels in $\mathcal{Y}$ 
\begin{equation}
  \label{eq1}
  \forall i \in \llbracket N - 1 \rrbracket, f_i : \mathbf{X}_i \longrightarrow \mathbf{y}_i, 
\end{equation}
and then \textit{(ii)} train another classifier $g$ to learn the mapping between the predicted labels $f_i(\mathbf{X}_N)$ from the pre-trained classifiers $f_i$,  and the true labels $\mathbf{y}_N$ for the remaining $N^{th}$ subject's features $\mathbf{X}_N$ (Fig.~\ref{figure:method}).
\begin{equation}
  \label{eq2}
  g : (f_1(\mathbf{X}_N), f_2(\mathbf{X}_N), ..., f_{N-1}(\mathbf{X}_N)) \longrightarrow \mathbf{y}_N
\end{equation}
We do this in each subject and report the average cross-validated accuracy.

\paragraph{Theory: different bias-variance decompositions.}
The interest of this ensembling can be understood by considering bias-variance decompositions. For simplicity, we consider Ridge regression and rely on known bounds \cite{hsu2014random}.
Dataset specific models $(f_i)_{i\in\llbracket N \rrbracket}$ yield a squared bias $\propto \sum_{j=1}^{N_{samples}} \frac{\lambda_j}{(1 + \lambda_j / \lambda)^2}(f_i^j)^2$, where  $f_i^j$ is the reponse magnitude to the $j^{th}$ dimension of the input, $(\lambda_j)$ are the eigenvalues of $\mathbf{X}_i$'s covariance matrix and $\lambda$ is the regularization parameter; the prediction variance is $\propto
\frac{var(\mathbf{y}) d(\lambda)}{N_{samples}}$, where $d(\lambda)=\sum_{j=1}^{N_{samples}} \left( \frac{\lambda_j}{\lambda_j + \lambda}\right)^2$ is the effective dimension of the Ridge model. For optimal $\lambda$, both terms are balanced, hence the squared prediction error is $\propto  \frac{var(\mathbf{y})d(\lambda)}{N_{samples}}$.
If the ensembling model $g$ is ordinary least squares, it has no estimation bias, but a model bias $\propto \frac{var(\mathbf{y})}{N}$ due to the projection of the feature space on the $(f_i(\mathbf{X}_N))_{i \in \llbracket N-1 \rrbracket}$ span, and a variance $\propto \frac{var(\mathbf{y})N}{N_{samples}}$. This leads to a squared error of order $\frac{var(\mathbf{y})}{N} + \frac{var(\mathbf{y})N}{N_{samples}}$.

As a consequence, we can outline three regimes: \textit{i)} if  $N$ is very small ($\mathcal{O}(1)$), then the ensemble model is affected by large modeling error  $\frac{var(\mathbf{y})}{N}$ and is not competitive;  \textit{ii)} if $N$ is larger, e.g. $\mathcal{O}(\sqrt{N_{samples}})$ then both error terms are balanced, making the ensemble model competitive with respect to the standard model $(N < d(\lambda))$ ; \textit{iii)} when $N$ becomes as large as $N_{samples}$, the ensemble model is dominated by variance; it thus requires regularization and may no longer be competitive. Overall, we can expect accuracy gains when $1 \ll N \ll N_{samples}$.

\paragraph{Datasets and tasks:}
We compare the ensemble approach against the conventional one in five different fMRI datasets and tasks with different characteristics (Table~\ref{table:datasets}).
Each of these datasets has $4$ to $6$ different conditions (or classes) of stimuli to be decoded.
All these datasets are publicly available and their preprocessing pipelines are described in detail in their respective publication.
We further use Nilearn \cite{Abraham2014} to extract brain signals, detrend and standardize these preprocessed data.
These datasets have varying voxel resolution, thus we down-sampled them to $3mm$ isotropic resolution for consistency, and to reduce computational costs.
Additionally, we applied a $5mm$ Gaussian smoothing kernel to improve the signal-to-noise ratio.
Finally, we fit a general linear model (GLM) to each dataset to derive trial-by-trial effect size maps using the Least Squares Separate approach \cite{Turner2012}.
These trial-by-trial GLM effect size maps are used as features for decoding.

\begin{table}[tb]
\centering
\caption{\textbf{List of fMRI datasets and corresponding tasks used in the study:} $N_{samples}$ refers to the number of samples per subject, $N_{subjects}$ to the number of subjects in the cohort and $N_{classes}$ to the number of classes in the prediction task.}
\label{table:datasets}
\begin{tabular}{|c|c|c|c|c|l|}
\hline
Dataset & Task & $N_{samples}$ & $N_{subjects}$ & $N_{classes}$ & Stimuli labels \\
\hline
Neuromod \cite{Bellec2024} & Visual n-back & 50 & 4 & 4 & images of body/face/ \\
& & & & & place/tools \\
AOMIC \cite{Snoek2021} & Emotion & 61 & 203 & 4 & negative/neutral \\
& anticipation \cite{Oosterwijk2017} & & & & emotion images and  \\
& & & & & cue for negative/neutral \\ 
& & & & & emotion images \\
Forrest \cite{Hanke2015} & Music genre & 175 & 10 & 5 & ambient/country/metal/ \\
& perception \cite{Casey2012} & & & & rocknroll/symphonic \\
& & & & & music \\
BOLD5000 \cite{Chang2019} & Image-Net \cite{Deng2009} & 332 & 3 & 4 & images of furniture/ \\ 
& image viewing & & & & vehicle/animal/person \\
RSVP-IBC \cite{Pinho2020} & RSVP & 360 & 13 & 6 & type of text: \\
& language \cite{Humphries2006} & & & & jabberwocky/complex/  \\
& & & & & simple/word list/ \\
& & & & & pseudoword list/ \\
& & & & & consonant strings \\
\hline
\end{tabular}
\end{table}
\paragraph{Decoding settings:}
In this work, we compare the two decoding approaches in two different feature spaces: the image voxel space (50k voxels) and a low-resolution space, based on the 1024-dimensional DiFuMo decomposition \cite{Dadi2020}. 
These DiFuMo features are obtained by regression using Nilearn \cite{Abraham2014}.
Furthermore, we also compare three different classifiers for decoding: Multi-layer Perceptron (MLP), linear Support Vector Classifier (SVC), and Random Forest \cite{Pedregosa2011}.
We use the default Scikit-learn parameters for each classifier with a few modifications.
For MLP, we use 100 hidden layers, ReLU activation, ADAM solver and 1000 iterations.
For SVC, we use $l2$ penalty, and square hinge loss, and the algorithm to solve dual optimization is set to automatic.
For Random Forest we use 500 trees, and for the maximum depth of a tree, the nodes are expanded until all leaves are pure or until all leaves contain less than 2 samples.
Note that only the final classifier ($g$ in Eq.~\ref{eq2}) is switched between the three model families, and the pre-trained classifiers ($f_i$ in Eq.~\ref{eq1}) are always linear SVC with $l2$ penalization as this remains the best-performing classifier for this type of data.
Results with $l1$ penalization for $f_i$ are given in Supplementary 
Fig. 2.

\noindent
\textit{Varying the training-set size:} Within each dataset, we keep 90\% of data for training and 10\% for testing. 
We vary the size of the training set over 10 geometrically increasing sub-samples of that initial 90\% training split and always test the trained model on the same 10\% testing split.
We do this for 20 different cross-validation train-test splits.
Note that in the ensemble approach, while pre-training the classifiers, we use all the samples available in each subject.

\noindent
\textit{Varying the number of subjects in the ensemble:} Next, we randomly sample a subset of subjects from each dataset and only use the pre-trained classifiers from these subjects to train the final classifier.
For each subset of subjects, we also vary the training set size as described in the previous paragraph.
We do this for 5 different cross-validation train-test splits, such that each split has a different subset of subjects, whenever possible.

\noindent
\textit{Evaluation metric:} We use the average (balanced) accuracy across all subjects in the dataset and twenty cross-validation splits, over ten training set sizes.
Balancing is done to account for the class imbalance for BOLD5000, where the metric used is the average recall across classes.
To get an estimate of the uncertainty in the average accuracy, we use a bootstrap approach as implemented in the Python library Seaborn.
The error bars represent a 95\% interval of the bootstrap distribution. 

\noindent
\textit{Code availability:}
All the code used in this study is available at: \\
\url{https://github.com/man-shu/ensemble-fmri}

\section{Results}
\label{section:results}
\begin{figure}[tb]
\centerline{\includegraphics[width=\linewidth]{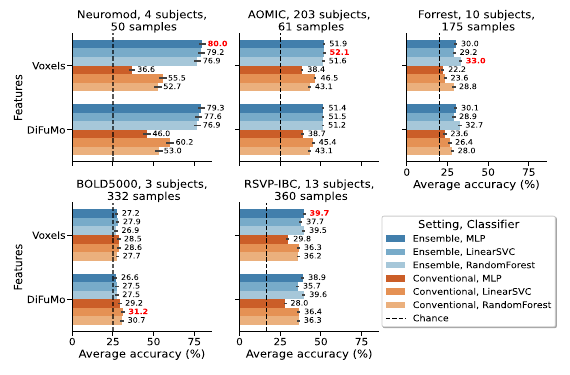}}
\caption{\textbf{Average decoding accuracy:} Each plot represents a different dataset (along columns). The average decoding accuracy is plotted along the x-axis. The averages are across all training sizes, subjects and 20 cross-validation splits. The error bars represent a 95\% confidence interval of the bootstrap distribution. The horizontal line represents the chance level of accuracy.}
\label{figure:average_accuracy}
\end{figure}

\paragraph{The Ensemble approach outperforms conventional decoding:}
Fig.~\ref{figure:average_accuracy} shows the average decoding accuracies for the conventional and ensemble approaches across all datasets and decoding settings.
The ensemble approach outperforms the conventional approach in all datasets and decoding settings, except BOLD5000.
The best average performance gain is observed in the Neuromod dataset with 20\% gains between best-performing conventional and ensemble approaches (Ensemble, Voxels, MLP vs. Conventional, DiFuMo, LinearSVC).
Using full-voxel feature space for pre-training is more beneficial than a reduced feature space, in all datasets, irrespective of the classifier used.
Comparisons between classifiers, however, are not straightforward.
In general, we see that MLP is the best-performing classifier --- at least it performs similarly to the best-performing classifier in all datasets and decoding settings.
However, in datasets where 10 or more subjects are available, and the number of samples is higher (as in Forrest and RSVP-IBC), Random Forests outperform linear SVC.
On the other hand, in datasets where the number of subjects is higher (as in AOMIC) or when data is scarce (as in Neuromod), linear SVC outperforms Random Forest.

In addition to the gain in accuracy, we could also extract feature importance scores for each dataset/task (Supplementary Fig. 1).
These scores help explore the spatial patterns of brain activity that are most informative for decoding and could hence be informative for understanding the cognitive processes.
For example, for the visual n-back task of Neuromod, the most informative voxels are located in the visual cortex; for the music genre perception task of Forrest, in the auditory cortex; and for the RSVP language task of RSVP-IBC, in the language processing areas.

\paragraph{The case of scarce data and many subjects:}
\begin{figure}[tb]
\centerline{\includegraphics[width=0.85\linewidth]{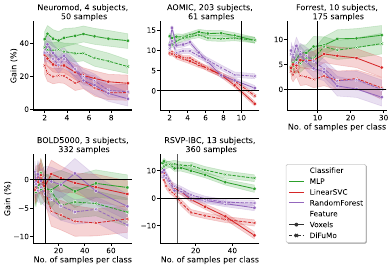}}
\caption{\textbf{Gain in decoding accuracy when varying the number of training samples per class:} Each plot represents a different dataset (along columns). The y-axis shows the average percent gain in decoding accuracy (accuracy of ensemble - accuracy of conventional) across all subjects and 20 cross-validation splits. On the x-axis, training size is reported as the number of samples per class in each cross-validation split. The confidence intervals represent 95\% confidence interval of bootstrap distribution. The horizontal line represents no average gain in accuracy and the vertical line, 10 samples per class.}
\label{figure:gain_v_samples}
\end{figure}
We further investigate the effect of the number of training samples per class on accuracy.
The \emph{gain in accuracy} displayed in Fig.~\ref{figure:gain_v_samples} is the difference in accuracy between the ensemble and conventional approaches.
In all datasets, with MLP as the final classifier, the gains are much higher relative to the other two classifiers.
This is due to the poor performance of MLP in the conventional approach in the scarce data regime.
On the other hand, with linear SVC and Random Forest, gains are observed when up to 10 samples per class are available for training in all datasets, except BOLD5000 (Fig.~\ref{figure:gain_v_samples}).
When more than 10 samples per class are available, the ensemble approach performs similarly to the conventional approach in all datasets, except BOLD5000 and RSVP-IBC.
Note that both datasets have the largest number of samples per class and this observation indicates that the ensemble approach is only beneficial when the number of samples per class is low.
%s
Furthermore, it should be noted that the BOLD5000 dataset has the lowest number of subjects (3) among all datasets and the loss of performance in this case shows that the availability of very few subjects hampers the ensemble approach even when the data is scarce.
Therefore, we also examined the effect of the number of subjects on the gain in accuracy.
\begin{figure}[tb]
\centerline{\includegraphics[width=.85\textwidth]{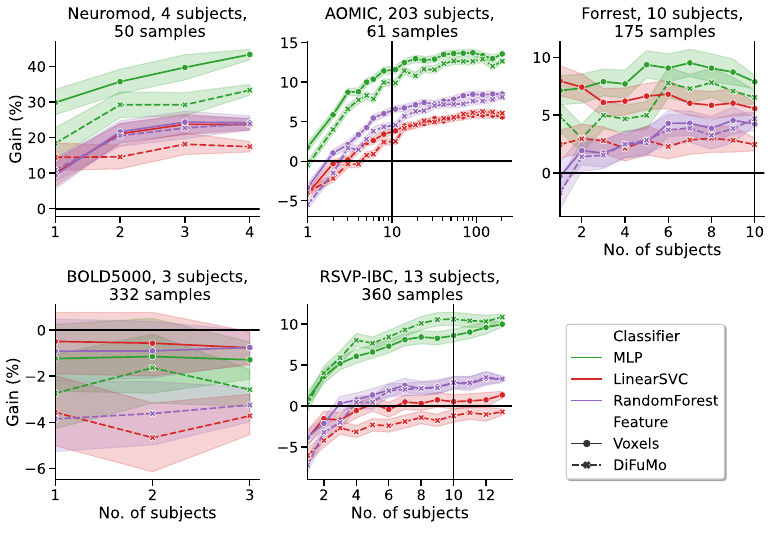}}
\caption{\textbf{Gain in decoding accuracy over a varying number of subjects in the ensemble:} Each plot represents a different dataset (along columns). The x-axis represents the number of subjects used in the ensemble method. The y-axis represents the average percent gain in decoding accuracy (accuracy of ensemble - accuracy of conventional) across all training sizes and 5 cross-validation splits. The confidence intervals represent 95\% interval of bootstrap distribution. The horizontal line represents no average gain in accuracy and the vertical line at 10 subjects in the ensemble.}
\label{figure:gain_v_subjects}
\end{figure}

Fig.~\ref{figure:gain_v_subjects} shows that gains in accuracy increase with an increasing number of subjects in the ensemble in all datasets and decoding settings.
Again, the gains are higher with MLP as the final classifier and are positive even when there is only one subject in the ensemble.
For the other two classifiers, the AOMIC dataset with the largest number of subjects (203) shows that the gains become positive starting from $3-4$ subjects and saturate at around 10 subjects.
The saturation is observed even with the MLP classifier.
The RSVP-IBC dataset with the largest number of samples (360) and stimulus classes (6) shows that when more samples are available and the classification task is more complex, the gains continue to increase even with more than 10 subjects in the ensemble.
Additionally, we tested the ensemble approach with linear SVC with $l1$ penalization during pre-training.
The average decoding accuracies are lower relative to $l2$ penalization, in all datasets, except AOMIC and RSVP-IBC (Supplementary Fig. 2) -- the two datasets with the largest number of subjects.
This indicates that the $l1$ penalization is beneficial when the number of subjects is large.

\section{Discussion}
In this work, we set out to investigate the potential of an ensemble approach towards decoding cognitive states from fMRI signals. 
This approach leverages patterns learned from an ensemble of other subjects.
We found that it outperforms the conventional decoding approach in four of the five different datasets considered.
Particularly, as predicted by bias/variance analysis, this approach leads to gains in accuracy when data is scarce, and the gains increase with an increasing number of subjects in the ensemble.
The only dataset where the ensemble approach did not outperform the conventional approach was BOLD5000, which has the lowest number of subjects (3) among all datasets.
On average, using the full-voxel space as features is more beneficial than a reduced feature space, in all datasets, irrespective of the classifier used.
Using MLP as the final classifier is a safe choice, as it is the best-performing classifier or at least performs similarly to the best-performing classifier in all datasets and decoding settings.
The relative performance of linear SVC and Random Forest is sensitive to the number of subjects and samples available for training.

Given that decoding is a central tool in cognitive mapping and clinical applications, the need for large sample sizes is a major bottleneck.
Therefore, the present study serves as a proof of concept that combining pre-trained classifiers from different individuals allows for efficient use of available information and alleviates the need for large sample sizes in decoding fMRI brain signals.
On average, we observe that due to the gain in accuracy with this approach, the number of samples required for training can be reduced by $5-10$ samples per class, depending on the dataset.
Moreover, given that this approach provides a gain in prediction accuracy in scarce data conditions, it can also be beneficial in real-time BCI applications \cite{Fede2020,Sorger2020} where initial data scarcity is a major issue.
Furthermore, this method could be particularly useful in basic cognitive research that uses task fMRI, where a given cognitive domain could not be decoded accurately due to a small sample size or low signal-to-noise ratio in conventional settings.

The primary limitation of the approach is that it cannot work when very few subjects are available -- as seen in the case of the BOLD5000 dataset.
However, there is some evidence that similar approaches are immune to cross-dataset covariate shifts \cite{Gu2022}.
Therefore, by combining classifiers pre-trained on a large set of subjects as presented in this work, one can leverage large-scale public datasets like Amsterdam Open MRI Collection (AOMIC) \cite{Snoek2021}.
%
% \textbf{XXX possible discussion on inductive biases here.}
%
The second limitation is that the gains with ensembling are only obtained in the case of scarce per-subject data.
In the present study, except for trying $l1$ and $l2$ penalizations for linear SVC classifiers during pre-training, we did not explore any other pre-training strategies.
Pre-training deep learning models instead could be beneficial when larger sample sizes are available.
Finally, it would also be interesting to investigate if the approach can benefit even further from functional alignment techniques \cite{Thual2022} that improve across-subject correspondence.

\begin{credits}
\subsubsection{\ackname}
Thanks to the reviewers for their constructive feedback. Thanks to our colleagues Ana Fernanda Ponce and Alexis Thual for all the discussions and comments. This work is supported by the KARAIB AI chair (ANR-20-CHIA-0025-01), the ANR-23-CE23-0016 project, the ANR-22-PESN-0012 France 2030 program, and the HORIZON-INFRA-2022-SERV-B-01 EBRAINS 2.0 infrastructure project.
\subsubsection{\discintname}
The authors declare no competing interests.
\end{credits}
\clearpage
\bibliographystyle{splncs04}
\bibliography{Paper-2040}
\end{document}

\title{Supplementary Material}
\subtitle{Across-subject ensemble-learning alleviates the need for large samples for fMRI decoding}
\titlerunning{Across-subject ensemble-learning for fMRI decoding}
\author{Himanshu Aggarwal\inst{1}\orcidID{0000-0003-2451-0470} \and
Liza Al-Shikhley\inst{1}\orcidID{0009-0002-0297-4231} \and
Bertrand Thirion\inst{1}\orcidID{0000-0001-5018-7895}}
%
% index{Aggarwal, Himanshu}
% index{Al-Shikhley, Liza}
% index{Thirion, Bertrand}
%
\authorrunning{H. Aggarwal et al.}
\institute{Inria, CEA, Universit{\'e} Paris-Saclay,  Palaiseau, 91120, France \\
\email{himanshu.aggarwal@inria.fr}\\
\email{liza.alshikhley@gmail.com}\\
\email{bertrand.thirion@inria.fr}\\
\url{https://team.inria.fr/mind/}\\
}
\maketitle
\begin{figure}
\centering{\includegraphics[width=\textwidth]{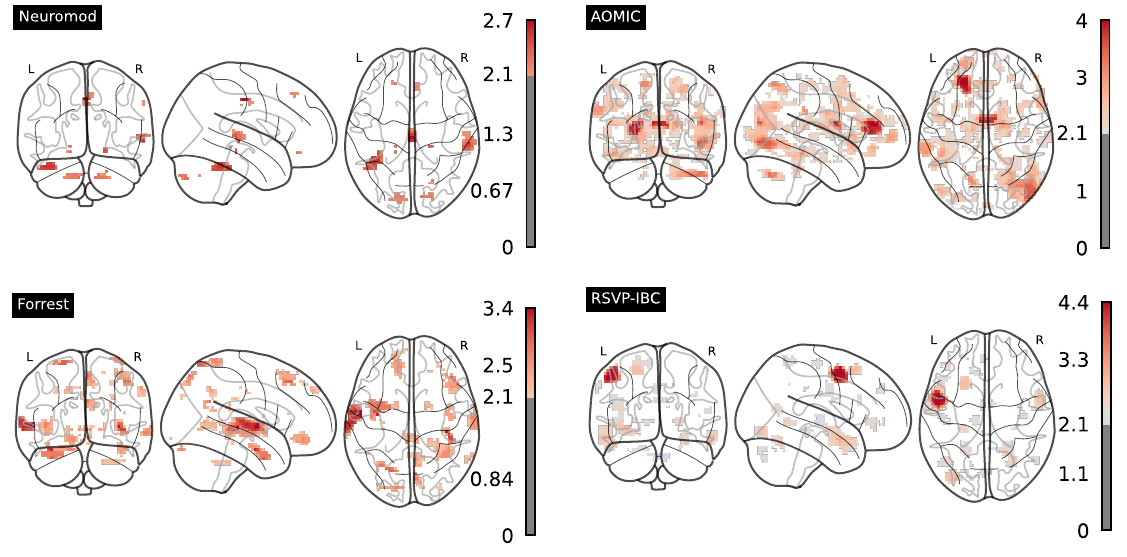}}
\caption{\textbf{Voxel-wise feature importance scores for one subject in each dataset:} The scores are z-scored and thresholded to only show the top 1 percentile. From left to right, top to bottom: the cognitive tasks performed in Neuromod is visual N-back, in AOMIC is emotion anticipation, in Forrest is music genre perception, and in RSVP-IBC is RSVP language task.
\label{figure:featimp_glass}
}

\end{figure}
\begin{figure}
\centering{\includegraphics[width=\textwidth]{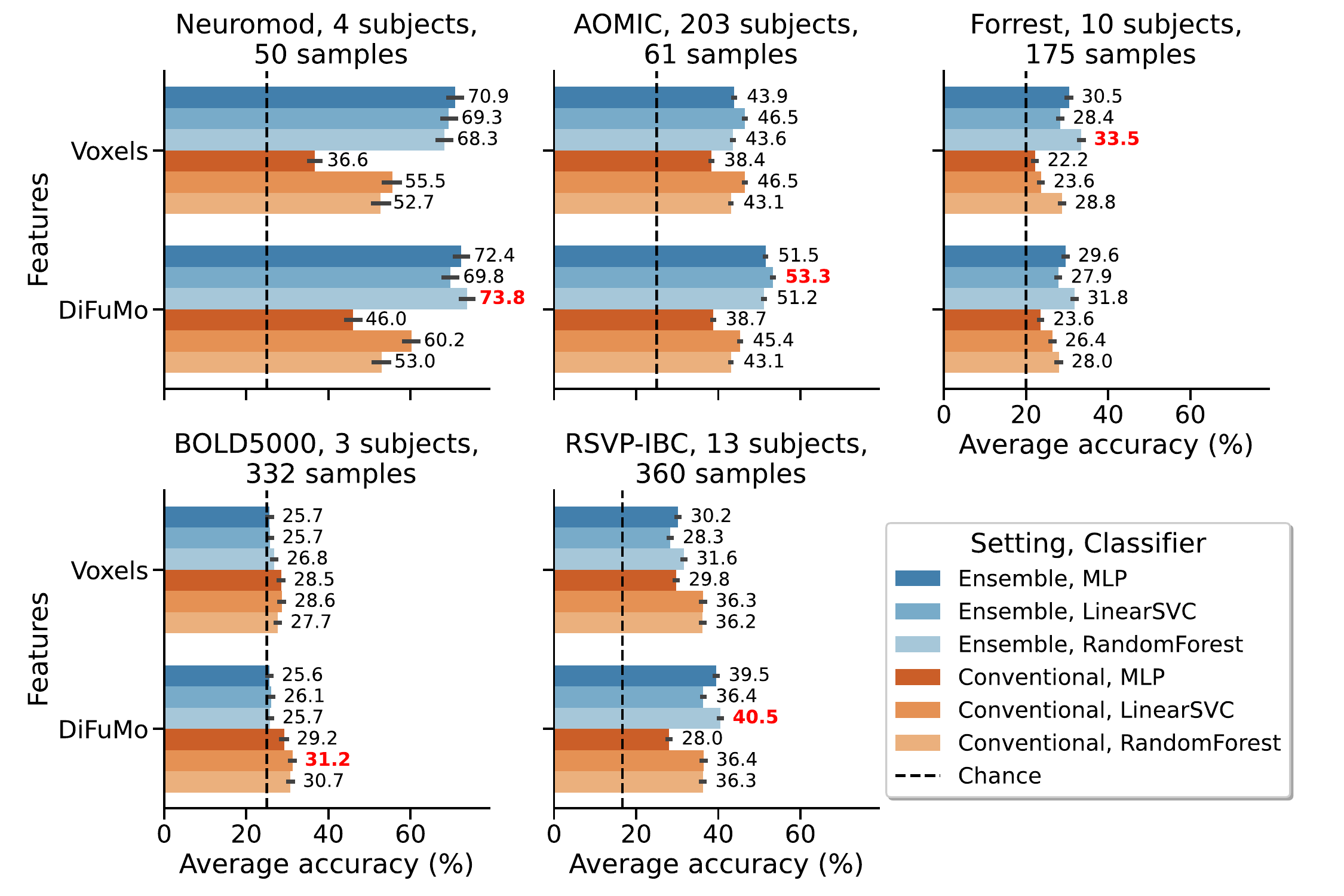}}
\caption{\textbf{Decoding accuracy using linear SVC with $l1$ penalization during pre-training:} Overall average decoding accuracy across subjects, 20 cross-validation splits and ten training set sizes.
\label{figure:accuracy_l1_penalty}
}
\end{figure}